\documentclass[journal]{IEEEtran}

\ifCLASSINFOpdf
\else
\fi
\usepackage{graphicx}          

\usepackage{ntheorem}
\usepackage[T1]{fontenc}

\usepackage[11pt]{moresize}
\usepackage[euler]{textgreek}
\usepackage{enumerate}
\usepackage[centertags]{amsmath}
\usepackage{graphicx}      
\usepackage{newlfont}
\usepackage{dsfont}
\usepackage{amssymb,amsmath,amsfonts}
\usepackage{epstopdf}
\usepackage[bf,normalsize,center]{subfigure}
\usepackage{multirow}

\usepackage{caption}
\usepackage{color}
\usepackage{cuted}

\makeatletter
\g@addto@macro\normalsize{%
  \setlength\abovedisplayskip{2pt}
  \setlength\belowdisplayskip{2pt}
}
\makeatother

\newcommand\norm[1]{\left\lVert#1\right\rVert}

\newcommand{\Real}{{\mathds R}} 
\newcommand{\Nat}{{\mathds N}} 

\newtheorem{definition}{Definition}{}
\newtheorem{corollary}{Corollary}{}
\newtheorem{proposition}{Proposition}{}
\newtheorem{theorem}{Theorem}{}
\newtheorem{remark}{Remark}{}
{}

\begin{document}

\title{Characterization of Model-Based Detectors for CPS Sensor Faults/Attacks}
%
%
%

\author{Carlos~Murguia
        and~Justin~Ruths
\thanks{Carlos Murguia is with the Center for Research in Cyber Security \textbf{(iTrust)}, Engineering Systems and Design (ESD) Pillar, Singapore University of Technology and Design, Singapore. E-mail: murguia\_rendon@sutd.edu.sg.}
\thanks{Justin Ruths is with the Departments of Mechanical and Systems Engineering, University of Texas at Dallas, USA, e-mail: jruths@utdallas.edu. }
\thanks{This work was supported by the National Research Foundation (NRF), Prime Minister's Office, Singapore, under its National Cybersecurity R\&D Programme (Award No. NRF2014NCR-NCR001-40) and administered by the National Cybersecurity R\&D Directorate.}
\thanks{Manuscript received May 5, 2017; revised May 5, 2017.}}

%
%

\markboth{Oct~2017}%
{Shell \MakeLowercase{\textit{et al.}}: Bare Demo of IEEEtran.cls for IEEE Journals}
%



\maketitle

\begin{abstract}
A vector-valued model-based cumulative sum (CUSUM) procedure is proposed for identifying faulty/falsified sensor measurements. First, given the system dynamics, we derive tools for tuning the CUSUM procedure in the fault/attack free case to fulfill a desired detection performance (in terms of false alarm rate). We use the widely-used chi-squared fault/attack detection procedure as a benchmark to compare the performance of the CUSUM. In particular, we characterize the state degradation that a class of attacks can induce to the system while enforcing that the detectors (CUSUM and chi-squared) do not raise alarms. In doing so, we find the upper bound of state degradation that is possible by an undetected attacker. We quantify the advantage of using a dynamic detector (CUSUM), which leverages the history of the state, over a static detector (chi-squared) which uses a single measurement at a time. Simulations of a chemical reactor with heat exchanger are presented to illustrate the performance of our tools.
\end{abstract}


\section{Introduction}

During the past half-century, scientific and technological advances have greatly improved the performance of control systems. From heating/cooling devices in our homes, to cruise-control in our cars, to robotics in manufacturing centers. However, these new technologies have also led to vulnerabilities of some our most critical infrastructures--e.g., power, water, transportation. Advances in communication and computing power have given rise to adversaries with enhanced and adaptive capabilities. Depending on attacker resources and system defenses, attackers may deteriorate the functionality of systems even while remaining undetected. Therefore, designing efficient fault/attack detection schemes and attack-robust control systems is of key importance for guaranteeing the safety and proper operation of critical systems. Tools from sequential analysis and fault detection have to be adapted to deal with the systematic, strategic, and persistent nature of attacks. These new challenges have attracted the attention of many researchers in the control and computer science communities \cite{Pasqualetti_1}\nocite{Mo_1}\nocite{Kwon}\nocite{Pappas}\nocite{Pappas}\nocite{Gupta}\nocite{Gupta2}\nocite{Cardenas}\nocite{Carlos_Justin2}\nocite{Carlos_Justin}-\cite{Carlos_Justin3}. Lately, there has been increasing interest in studying \emph{systems performance degradation} induced by attacks that remain hidden or undetected by detection procedures \cite{Pasqualetti_1}-\cite{Mo_1},\cite{Gupta}-\cite{Gupta2},\cite{Carlos_Justin3}. Quantifying the system degradation provides \emph{a measure of impact} to assess the performance of control structures, estimation schemes, and detection procedures against this class of intelligent attacks. For instance, in \cite{Gupta}-\cite{Gupta2}, for arbitrary detection procedures, the authors quantify how much the attacker can deviate the estimate of the state from its attack-free values while remaining stealthy. They characterize stealthiness of attacked sequences using the \emph{Kullback-Leibler Divergence} \cite{Ross} between the attack-free and the attacked sequence. In the same spirit, the authors in \cite{Pasqualetti_1}-\cite{Mo_1} study how attacks propagate through the control structure to degrade the system dynamics while remaining \textit{undetected} by the detection mechanism. In particular, the authors in \cite{Pasqualetti_1} characterize undetectability (for a class of deterministic LTI systems) as the ability of attackers to excite \emph{only} the zero dynamics of the system \cite{Henk2} (making its effect undetectable from output measurements). In \cite{Mo_1},\cite{Carlos_Justin3},\cite{guo2017optimal}, the authors propose a notion of stealthiness by attacks that do not change the alarm rate of the detector by more than a small amount (thus making it hard for the operator to distinguish between an attack-free and an attacked system, i.e., these attacks remain \textit{hidden} from the detector). As a measure of impact, they characterize the reachable sets that these hidden attacks can induce to the system.

Most of the current work on security of control systems has focused on \textit{static} detection procedures (either bad-data or chi-squared detectors), which identify anomalies based on a single measurement at a time \cite{Pasqualetti_1}\nocite{Mo_1}\nocite{Kwon}-\cite{Pappas},\cite{Carlos_Justin3}. There is only a small amount of literature considering the use of dynamic change detection procedures such as the Sequential Probability Ratio Test (SPRT) or the Cumulative Sum (CUSUM) \cite{Gustafsson}, which employ measurement history, in the context of security of Cyber-Physical Systems (CPS) \cite{Cardenas},\cite{Kwon22}-\cite{Urbina:2016:LIS:2976749.2978388}. Dynamic detectors present an appealing alternative to the aforementioned static procedures. Using measurement history provides extra degrees of freedom for improving the performance of our fault/attack detection strategies; in particular, against low amplitude persistent attacks \cite{Urbina:2016:LIS:2976749.2978388}.

This paper addresses, for Linear Time-Invariant (LTI) systems subject to sensor/actuator noise, the problem of characterizing CUSUM dynamic and chi-squared static detectors in terms of false alarm rates and performance degradation under a class of attacks. Standard Kalman filters are proposed to estimate the state of the physical process. Both detectors employ a distance measure that is a quadratic function of the residual (the error between sensor measurements and the estimated outputs). In the chi-squared procedure, at each time instant if the distance measure is larger than a threshold, an alarm is raised, indicating a possible compromised sensor. In the CUSUM procedure, the distance measure values are accumulated over time, and if this accumulated value is greater than expected an alarm is triggered. Fundamentally a detector aims to properly raise an alarm when a fault/attack happens and not raise an alarm when there is no fault/attack. Deviation from this ideal performance is captured by false positives (alarms are raised when there are no faults/attacks), also called false alarms, and false negatives (a fault/attack happens, but no alarm is raised). Although minimizing both false positives and false negatives is best, often they must be traded off based on which is more tolerable. In this context, detectors with high sensitivity would have high rates of false positives in favor of low rates of false negatives (and vice versa).

In order to provide an equitable comparison between detectors (e.g., static versus dynamic), we first require the ability to tune each type of detector to a similar level of sensitivity. To-date, however, there is not a complete characterization of how features of the system (e.g., system matrices, control/estimator gains, noise, sampling) affect the selection of the CUSUM parameters to achieve a desired sensitivity, quantified by the rate of false alarms. Our first contribution in this paper is to provide systematic tools to tune the CUSUM detector, and for completeness, also the chi-squared (static) detector, in the fault/attack free case based on the system dynamics, the Kalman filter, the stochastic properties of the distance measure, and a desired false alarm rate. In particular, sufficient conditions for mean square boundedness of the CUSUM sequence are derived when it is driven by a quadratic form of the residual. Then, using a Markov chain approximation of the CUSUM sequence, we give a procedure for selecting the \emph{decision threshold} such that a desired false alarm rate is satisfied.

Second, for a class of \emph{zero-alarm attacks} (attacks that prevent the detector from raising alarms), we characterize the impact of the attack sequence on the system dynamics when the vector-valued CUSUM and chi-squared detectors are deployed for attack detection.  From an empirical point of view, zero-alarm attacks have been actively used to assess the resilience of dynamical systems against attacks \cite{Cardenas},\cite{Urbina:2016:LIS:2976749.2978388},\cite{Urbina2016SurveyAN}. zero-alarm attacks provide a simple, deterministic, yet representative class of attacks that can be easily scaled to systems with different dynamics and properties; thus making them a good choice for assessing the performance of attack detectors in terms of state degradation.

In our preliminary work \cite{Carlos_Justin2}, we have started analyzing these ideas. The contributions of this manuscript with respect to \cite{Carlos_Justin2} are the following: A comprehensive and complete exposition of all the results and methodologies; the main results have been revised and improved and the corresponding proofs (which are not given in our preliminary work) are included in this paper; we formulate an all-new measure for attack degradation centered around the concept of input-to-state stability; and a benchmark simulation experiment used in the fault-detection literature \cite{Patton_Book},\cite{Wata} (a chemical reactor with heat exchanger) is presented to illustrate the performance of our tools.

\subsection{Notation}
Throughout this paper, the following notation is used: the symbol $\Real$ stands for the real numbers, $\Real_{>0}$($\Real_{\geq 0}$) denotes the set of positive (non-negative) real numbers. The symbol $\Nat$ stands for the set of natural numbers. The Euclidian norm in $\Real^n$ is denoted by $\norm{x}$, $\norm{x}^2=x^Tx$, where $^T$ denotes transposition. The induced norm of a matrix $A \in \Real^{n \times n}$, denoted by $\norm{A}$, is defined as $\norm{A} = \max_{x \in \Real^n,\norm{x}=1} \norm{Ax}$. The $n \times n$ identity matrix is denoted by $I_n$ or simply $I$ if no confusion can arise. Similarly, $n \times m$ matrices composed of only ones and only zeros are denoted by $\mathbf{1}_{n \times m}$ and $\mathbf{0}_{n \times m}$, respectively, or simply $\mathbf{1}$ and $\mathbf{0}$ when their dimensions are clear. If a quadratic form $x^TPx$ with a symmetric matrix $P=P^T$ is positive definite (semidefinite), then $P$ is called positive definite (semidefinite). For positive definite (semidefinite) matrices, we use the notation $P>0$ ($P \geq 0$); moreover, $P>Q$ ($P \geq Q$) means that the matrix $P-Q$ is positive definite (semidefinite). The spectrum of a matrix $A$ is denoted by $\text{spec}[A]$, $\text{tr}[A]$ denotes its trace, and $\rho[A]$ is its spectral radius.
The notation $\lambda_{\min}[A]$ ($\lambda_{\max}[A]$) stands for the smallest (largest) eigenvalue of the square matrix $A$. The notation $E[x]$ stands for the expected value of $x$ and $E_y[x]$ denotes the expected value of $x$ conditional to $y$. The variance of a random variable $x$ is denoted by $\text{var}[x]$. The notation $\text{pr}[\cdot]$ denotes probability and $x \sim \mathcal{N}(\mu,\Sigma)$ means that $x \in \Real^n$ is a vector-valued normally distributed random variable with mean $\mu \in \Real^n$ and covariance matrix $\Sigma \in \Real^{n \times n}$. For simplicity of notation, we often suppress the explicit dependence of time $t$.

\section{System Description \& Attack Detection}

\begin{figure}
  \centering
  \includegraphics[scale=.3]{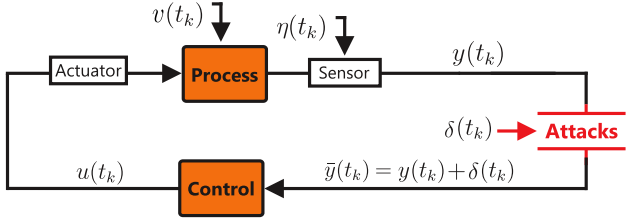}
  \caption{Cyber-physical system under sensor attacks.}\label{Fig1}
\end{figure}

We study LTI stochastic systems of the form:
\begin{equation}
\left\{
\begin{array}{ll}
{x}(t_{k+1}) = Fx(t_k) + G u(t_k) + v(t_k),  \label{1}\\
y(t_k) = Cx(t_k) + \eta(t_k),
\end{array}
\right.
\end{equation}
with sampling time-instants $t_k,k \in \Nat$,  state $x \in \Real^n$, measured output $y \in \Real^m$, control input $u \in \Real^l$, matrices $F$, $G$, and $C$ of appropriate dimensions, and i.i.d.  multivariate zero-mean Gaussian noises $v \in \Real^n$ and $\eta \in \Real^m$ with covariance matrices $R_{1} \in \Real^{n \times n}$, $R_1 \geq 0$ and $R_2 \in \Real^{m \times m}$, $R_2 \geq 0$, respectively. The initial state $x(t_1)$ is assumed to be a Gaussian random vector with covariance matrix $R_0 \in \Real^{n \times n}$, $R_0 \geq 0$. The processes $v(t_k)$, $k \in \Nat$ and $\eta(t_k)$, $k \in \Nat$ and the initial condition $x(t_1)$ are mutually independent. It is assumed that $(F,G)$ is stabilizable and $(F,C)$ is detectable. At the time-instants $t_k,k \in \Nat$, the output of the process $y(t_k)$ is sampled and transmitted over a communication network. The received output $\bar{y}(t_k)$ is used to compute control actions $u(t_k)$ which are sent back to the process, see Fig. \ref{Fig1}. The complete control-loop is assumed to be performed instantaneously, i.e., the sampling, transmission, and arrival time-instants are equal. In this paper, we focus on attacks on sensor measurements. That is, in between transmission and reception of sensor data, an attacker may replace the signals coming from the sensors to the controller\footnote{Such an attack can also be accomplished by installing malware on the controller equipment, in which case true measurements reach the controller, but are manipulated before they are used.}, see Fig. \ref{Fig1}. After each transmission and reception, the attacked output $\bar{y}$ takes the form:
\begin{equation}
  \bar{y}(t_k) := y(t_k) + \delta(t_k) = Cx(t_k) + \eta(t_k) + \delta(t_k), \label{3}
\end{equation}
where $\delta(t_k) \in \Real^m$ denotes \emph{additive sensor attacks/faults}. Denote $x_k:=x(t_k)$, $u_k:= u(t_k)$, $v_k:=v(t_k)$, $\bar{y}_k:=\bar{y}(t_k)$, $\eta_k:=\eta(t_k)$, and $\delta_k:=\delta(t_k)$. Using this new notation, the attacked system is written in the following compact form:\linebreak
\begin{equation}
\left\{
\begin{array}{ll}
{x}_{k+1} = F x_k + G u_k + v_k,\label{17} \\
\text{ \ \ }\hspace{.8mm}\bar{y}_k = C x_k + \eta_k + \delta_k.
\end{array}
\right.
\end{equation}

\begin{remark}
If the stochastic processes $v_c(t_k)$ and $\eta(t_k)$ are non-Gaussian, using spectral factorization \emph{\cite{Tri,Ljung}}, we could rewrite them as output signals coming from linear\- filters, say $G_1(q)$ and $G_2(q)$, with Gaussian stochastic processes as inputs, say $w_1(t_k)$ and $w_2(t_k)$; that is, $v_c(t_k) = G_1(q)w_1(t_k)$ and  $\eta(t_k) = G_2(q)w_2(t_k)$, where $q$ denotes the forward-shift operator. Then, by extending the system dynamics with the filters and considering the non-Gaussian noises $v_c(t)$ and $\eta(t_k)$ as new states, the extended system is written\- as a LTI system perturbed by Gaussian noise, see, for instance, \emph{\cite{Tri,Ljung}} for details.
\end{remark}

\subsection{Steady state Kalman filter (attack/fault free case)}

To estimate the state of the process, a one step-ahead estimator with the following structure is proposed:
\begin{equation}
\hat{x}_{k+1} = F \hat{x}_k + Gu_k + L_k\big( \bar{y}_k - C\hat{x}_k   \big),  \label{19}
\end{equation}
with estimated state $\hat{x}_k \in \Real^n$, $\hat{x}_1 = E[x(t_1)]$, and gain \linebreak matrix $L_k \in \Real^{n \times m}$. Define the estimation error $e_k:= x_k - \hat{x}_k$. The matrix $L_k$ is designed to minimize the covariance matrix $P_k:= E[e_ke_k^T]$ in the absence of attacks. Given the discrete-time dynamics (\ref{17}) and the estimator (\ref{19}), the estimation error is governed by the difference equation:
\begin{align}
e_{k+1} = \big( F - L_kC \big) e_k - L_k\eta_k - L_k \delta_k + v_k. \label{20}
\end{align}
If the pair $(F,C)$ is detectable, the covariance matrix converges to steady state in the sense that, in the attack-free case, $\lim_{k \rightarrow \infty} P_k = P$ exists \cite{Astrom}. Let $\delta_k=\mathbf{0}$; then, from (\ref{20}), the mean value of $e_k$ is given by
\begin{align}
E[e_{k+1}] = \big( F - L_kC \big) E[e_k]. \label{20a}
\end{align}
Because $\hat{x}_1=E[x(t_1)]$, the mean value of the estimation error equals $\mathbf{0}_{n \times 1}$ independent of $L_k$. We assume that the system has reached steady state before an attack occurs. Then, the estimation of the random sequence $x_k, k \in \Nat$ can be obtained by the estimator (\ref{19}) with $P_k$ and $L_k$ in steady state. It can be verified that, if $CPC^T + R_2$ is positive definite (a standard assumption that guarantees that the Kalman filter converges), the estimator gain:
\begin{align}
L_k = L := \big( FPC^T  \big)\big( R_2 + C P C^T  \big)^{-1}, \label{23}
\end{align}
leads to the minimal steady state covariance matrix $P$, with $P$ given by the solution of the algebraic Riccati equation:
\begin{equation}
F P F^T - P + R_1= FPC^T  ( R_2 + CPC^T  )^{-1} CPF^T.  \label{24}
\end{equation}
The reconstruction method given by (\ref{19})-(\ref{24}) is referred to as the \emph{steady state Kalman filter}, cf. \cite{Astrom}.\\

\begin{remark}
It is well known that, if the noise sequences $v_k$ and $\eta_k$ are Gaussian, the Kalman filter \eqref{19}-\eqref{24} gives the best estimate $\hat{x}_{k+1}$ of the state $x_{k+1}$ (in terms of minimum-mean-square estimation error) from noisy measurements. Moreover, if the noise is not Gaussian, the Kalman filter is the best linear estimator; although there may exist non\-linear estimators with better performance, cf. \emph{\cite{anderson}}.
\end{remark}

\subsection{Residuals and hypothesis testing}

Attacks can be regarded as  \emph{induced faults} in the system. Then, it is reasonable to use existing fault  detection techniques to identify sensor attacks. The main idea behind fault detection theory is the use of an \emph{estimator} to forecast the evolution of the system in the absence of faults. This prediction is compared with the actual measurements from the sensors. If the difference between what it is measured and the estimation (often referred to as \emph{residual}) is larger than expected, there might be a fault in the system. Although the notion of resi\-duals and model-based detectors is now routine in the fault detection literature, the primary focus has been on detecting\- and isolating faults with \emph{specific structures} (e.g., constant biases in sensor measurements or random faults in sensors and actuators following specific distributions). Now, in the context of an \linebreak intelligent adversarial attacker, new challenges arise to understand the effect that an intruder can have on the system given the dynamics, the estimator, and the detector structure. In this work, we use the steady state Kalman filter introduced in the previous section as our estimator.

Consider the discrete-time process dynamics (\ref{17}), the steady state Kalman filter (\ref{19})-(\ref{24}), and the corresponding error difference equation (\ref{20}). Define the residual sequence $r_k, k\in \Nat$ as
\begin{align}
r_k := \bar{y}_k - C\hat{x}_k = Ce_k + \eta_k + \delta_k. \label{25}
\end{align}
Then, $r_k$ evolves according to the difference equation:
\begin{equation}
\left\{
\begin{array}{ll}
e_{k+1} = \big( F - LC \big) e_k - L\eta_k + v_k - L \delta_k,  \label{26} \\
\text{ \ \ } r_k = Ce_k + \eta_k + \delta_k.
\end{array}
\right.
\end{equation}
If there are no faults/attacks, the mean of the residual is
\begin{equation}
E[r_{k+1}] = CE[e_{k+1}] + E[\eta_{k+1}] = \mathbf{0}_{m \times 1},  \label{27} \\
\end{equation}
and the covariance matrix is given by
\begin{align}
E[r_{k+1}r_{k+1}^T] &= CPC^T + R_2 =: \Sigma  \in \Real^{m \times m}.\label{28}
\end{align}
For this residual, we identify two hypothesis to be tested: $\mathcal{H}_0$ the \emph{normal mode} (no faults/attacks) and $\mathcal{H}_1$ the \emph{faulty mode} (with faults/attacks). Under the normal mode, the statistics of the residual are:
\begin{equation}
\mathcal{H}_0: \left\{
\begin{array}{ll}
E[r_k] = \mathbf{0}_{m \times 1},  \label{29} \\
E[r_kr_k^T] = \Sigma.
\end{array}
\right.
\end{equation}
Therefore, when an fault/attack occurs in the system ($\mathcal{H}_1$), we expect that the statistics of the residual are different from the normal mode, i.e.,
\begin{equation}
\mathcal{H}_1: \left\{
\begin{array}{ll}
E[r_k] \neq \mathbf{0}_{m \times 1},\text{ or}  \label{30} \\
E[r_kr_k^T] \neq \Sigma .
\end{array}
\right.
\end{equation}\\
There exist many well-known hypothesis testing techniques which may be used to examine the residual and subsequently detect faults/attacks. For instance, Sequential Probability Ratio Testing (SPRT) \cite{wald,Willsky}, Cumulative Sum (CUSUM) \cite{Gustafsson,Page}, Generalized Likelihood Ratio (GLR) testing \cite{Basseville}, Compound Scalar Testing (CST) \cite{Gertler}, etc. Each of these techniques has its own advantages and disadvantages depending on the scenario. The most utilized and powerful one is, arguably the SPRT, which minimizes the time to reach a decision for given probabilities of false detection (i.e., declaring $\mathcal{H}_1$ when it is actually $\mathcal{H}_0$). In this manuscript, we mainly focus on the CUSUM procedure which is a version of SPRT that permits repeated detection \cite{Page}. However, for comparison, we also present results about a particular case of CST, namely the so-called \emph{chi-squared} change detection procedure.

\subsection{Distance measures and CUSUM procedure}

Change detection theory was founded by Wald in 1947 when his book "\emph{Sequential Analysis}" was published and the SPRT was first introduced. Subsequently, the CUSUM procedure \cite{Page} was proposed by Page to detect changes in the mean of random variables by testing a weighted sum of the last few observations, i.e., a moving average. As Page pointed out, the CUSUM is equivalent to a repeated SPRT in which the test is restarted once a change has been detected. The input to the CUSUM procedure is a \emph{distance measure} $z_k \in \Real$, $k \in \Nat$, i.e., a measure of how deviated the estimator is from the sensor measurements. We propose the quadratic distance measure
\begin{equation}
z_{k} := r_{k}^T \Sigma^{-1} r_k,\label{32}
\end{equation}
where $r_k$ and $\Sigma$ are the residual sequence and its covariance matrix defined in (\ref{25}) and (\ref{28}), respectively. If there are no attacks,  $E[r_{k}] = \mathbf{0}$ and $E[r_kr_{k}^T] = \Sigma$; it follows that\\
\begin{equation}
\left\{
\begin{array}{ll}
\hspace{1.5mm} E[z_k]  &= \text{tr}[\Sigma^{-1} \Sigma] + E[r_k]^T \Sigma^{-1} E[r_k]  \\ &= m , \\
\text{var}[z_{k}] &= 2\text{tr}[\Sigma^{-1} \Sigma \Sigma^{-1} \Sigma] + 4 E[r_k]^T \Sigma^{-1}\Sigma\Sigma^{-1} E[r_k] \\ &= 2m ,
\end{array}
\right.\label{38}
\end{equation}
see, e.g., \cite{Ross} for details. Moreover, since $r_{k} \sim \mathcal{N}(\mathbf{0},\Sigma)$, then $z_{k} = r_{k}^T \Sigma^{-1} r_k$ follows a chi-squared distribution with $m$ degrees of freedom, cf. \cite{Ross}. Other options are based on \emph{likelihood ratios}. In this case, instead of directly using the sequence $z_k$ to drive the CUSUM procedure, the \emph{log-likelihood ratio} $\Lambda_k(z_k)$ between the two hypotheses is employed:
\begin{equation}
\Lambda_k(z_k) := \log\frac{f^1_{z_k}(z|\mathcal{H}_1)}{f^0_{z_k}(z|\mathcal{H}_0)},\label{34}
\end{equation}
where $f^j_{z_k}(z|\mathcal{H}_j)$ denotes the Probability Density Function (PDF) of the distance measure $z_k$, $k \in \Nat$ under $\mathcal{H}_j$, $j=\{0,1\}$. A problem to address when using log-likelihood ratios for detecting attacks or unstructured faults is the fact that the PDF of the faulty sequence $f_1(z_k|\mathcal{H}_1)$ is unknown. Actually, in the case of attacks, the adversary may induce any arbitrary (and possibly) non-stationary sequence $z_k$. Assuming the statistical properties of the attack sequences may limit our ability to detect a wide range of attacks \cite{Cardenas}.

The CUSUM procedure of Page driven by the distance measure $z_k$ is defined as follows.

\noindent\rule{\hsize}{1pt}\vspace{.5mm}
\textbf{CUSUM:} \vspace{0.5mm}
\begin{equation}
\left\{
\begin{array}{ll}
S_{1} = 0,\\
S_{k} = \max(0,S_{k-1} + z_{k} - b ), \hspace{.5mm} \text{ if } S_{k-1} \leq \tau, \label{35}   \\
S_{k} = 0 \text{ \ and \ } \tilde{k} = k-1,\ \ \hspace{.7mm}\text{ \ \ \ \hspace{.75mm}if } S_{k-1}  > \tau.
\end{array}
\right.
\end{equation}
\textbf{Design parameters:} bias $b \in \Real_{>0}$ and threshold $\tau \in \Real_{>0}$.\\
\textbf{Output:} alarm time(s) $\tilde{k}$.\\
\noindent\rule{\hsize}{1pt}

The idea is that the test sequence $S_{k}$ accumulates the distance measure $z_{k}$ and alarms are triggered when $S_{k}$ exceeds the threshold $\tau$. The test is reset to zero each time $S_{k}$ becomes negative or larger than $\tau$. If $z_{k}$ is an independent non-negative sequence (which is our case) and $b$ is not sufficiently large, the CUSUM sequence $S_{k}$ grows unbounded until the threshold $\tau$ is reached, no matter how large $\tau$ is set. In order to prevent these drifts, inevitably leading to false alarms, the bias $b$ must be selected properly based on the statistical properties of the distance measure. Once the the bias is chosen, the threshold $\tau$ must be selected to fulfill a required false alarm rate $\mathcal{A}^*$ (see Section \ref{False_alarma}).

\section{CUSUM-Tuning}

To enhance the performance of the CUSUM procedure, the bias $b$ and the threshold $\tau$ must be selected appropriately. We have already mentioned that too small a bias can lead to inevitable growth of the CUSUM test sequence. At the same time, too large a bias may hide the effect of faults/attacks. In what follows, we provide tools for selecting these parameters given the statistical properties of the distance measure $z_{k}$ introduced in \eqref{38}. In particular, we provide sufficient conditions on the bias $b$ such that, \emph{in the absence of faults/attacks}, the sequence $S_{k}$ of the CUSUM remains bounded (independent of the reset due to $\tau$) in mean-squared sense. This is important to avoid false alarms due to the inherent divergence of $S_{k}$. Subsequently, we characterize the {false alarm rate} of the CUSUM in terms of $b$ and $\tau$ given a \emph{desired} false alarm rate.

\subsection{Boundedness}

First, we introduce the following concept of boundedness of stochastic processes, cf. \cite{Agniel},\cite{Tarn}, followed by sufficient conditions for boundedness of the CUSUM sequence.\\

\begin{definition}
The sequence $S_{k}$, $k \in \Nat$ is said to be bounded in mean square, if
\[
\sup_{k \in \Nat} E_{S_{1}} \big[ S_{k}^2 \hspace{.25mm}\big] < \infty,
\]
is satisfied, i.e., the second moment of $S_{k}$ is finite.
\end{definition}

 \begin{theorem}\label{prop1}
Consider the discrete-time process \eqref{17} and the steady state Kalman filter \eqref{19}-\eqref{24}. Assume that there are no attacks to the system, i.e., $\delta_k = \mathbf{0}$. Let the \emph{CUSUM} \eqref{35} with bias $b \in \Real_{>0}$ and threshold $\tau \in \Real_{>0}$ be driven by the distance measure $z_{k} = r_{k}^T \Sigma^{-1} r_k$, $k \in \Nat$ with residual sequence $r_{k} \sim \mathcal{N}(\mathbf{0},\Sigma)$, $k \in \Nat$. Then, if the bias is set larger than the number of measurements, $b > \bar{b}:= m$, the \emph{CUSUM} sequence $S_{k}$, $k \in \Nat$ is bounded in mean square sense independent of the threshold $\tau$.\\

\emph{The proof of Theorem 1 is presented in the appendix.}\\
 \end{theorem}

\begin{remark}
Notice that boundedness of the first moment follows\- from boundedness of the second moment and Jensen's inequality \emph{\cite{Ross}}. Then, $b>\bar{b}=m$ implies that the expected\- value $E_{S_{1}}[S_{{k}}]$, $k \in \Nat$ is finite.\\
\end{remark}

The result stated in Theorem \ref{prop1} implies that for $b>\bar{b}$, the second\- moment (and hence the first) of the sequence $S_{k}$, $k\in \Nat$ does not diverge. Consequently, we avoid false alarms due to intrinsic growth of the CUSUM sequence. Note that if the bias $b$ is selected greater than but close to $\bar{b}$, small changes in the distance measure $z_{k}$ would lead to divergence of $S_{k}$. Therefore, the smaller the bias, the higher the sensitivity against changes in (or uncertain characterization of) the residual signals.\\

\subsection{False Alarms}\label{False_alarma}

Once the bias is selected such that boundedness of the se\-cond moment $E[S_{k}^2]$ is guaranteed, the next step is to select the thres\-hold $\tau$ to fulfill a desired false alarm rate. The occu\-rrence of an alarm in the CUSUM when there are no faults/attacks to the CPS is referred to as a false alarm. Operators need to tune this false alarm rate depending on the application. To do this, the threshold $\tau$ must be selected to fulfill a \emph{desired false alarm rate} $\mathcal{A}^*$. Let $\mathcal{A} \in [0,1]$ denote the \emph{false alarm rate} of the procedure defined as the expected proportion of observations which are false alarms, i.e., for the CUSUM procedure, $\mathcal{A}:=\text{pr}[S_k \geq \tau]$, see \cite{Dobben} and \cite{Adams}. Define the \emph{run length} $\mathcal{K}$ of the CUSUM (\ref{35}) as the number of ite\-rations needed such that $S_{\mathcal{K}} > \tau$ (without attacks):
\begin{equation}\label{65}
\mathcal{K} := \text{min}\{k\geq 1: S_{k} > \tau\}.
\end{equation}
The expected value $E[\mathcal{K}]$ of $\mathcal{K}$ is known in the literature as the \emph{Average Run Length} ($\text{ARL}$). The $\text{ARL}$ is inversely proportional to the false alarm rate $\mathcal{A}$ \cite{Adams,Dobben}, i.e.,
\begin{equation}\label{66}
\mathcal{A} = 1/\text{ARL}.
\end{equation}
Then, for a given $b>\bar{b}$, the problem of selecting $\tau$ to satisfy a desired false alarm rate $\mathcal{A}^*$ can be reformulated as the problem of selecting $\tau$ such that
\begin{equation}\label{67}
{\text{ARL}}= 1/\mathcal{A}^*.
\end{equation}
To determine a pair $(b,\tau)$ satisfying (\ref{67}), an expression for the $\text{ARL} = E[\mathcal{K}]$ is required but, in general, its exact evaluation is analytically intractable \cite{Khan}. The problem of approximating the ARL for CUSUM procedures has been addressed by many authors during the last decades. For instance, the authors in \cite{Khan}\nocite{Park}-\cite{Reynolds} propose Wiener process approximations of the ARL using analogies between the CUSUM and the SPRT for normally distributed distance measures. Although\- these techniques lead to explicit formulas for evaluating the ARL, the obtained approximations are often too conservative, see \cite{Park}-\cite{Reynolds}. Accurate numerical methods have been proposed by, for instance, \cite{Champ}\nocite{Evans}\nocite{Luceno}-\cite{Woodall}. These methods rely on two main techniques, namely Markov chain and integral equation approaches. Both methods give accurate predictions of the ARL  (see \cite{Champ} for a comparison); however, we find the Markov chain approach more constructive and easier\- to implement. In this work, we use the result of Evans and Brook \cite{Evans}. With this result, we outline a procedure for selecting the threshold $\tau$ given the bias $b$ and a required false alarm rate $\mathcal{A}^*$.

For given $b>\bar{b}$ and some $\tau \in \Real_{>0}$, consider the sequence $S_{k}$ generated by the CUSUM procedure (\ref{35}) driven by the distance measure $z_{k} = r_{k}^T \Sigma^{-1} r_k$, $k \in \Nat$. Given the recursive nature of the CUSUM procedure and independence of $v_{k}$ and $\eta_{k}$, $k \in \Nat$, the sequence $S_{k}$ forms a Markov chain taking\- values on the non-negative real line  \cite{Meyn}. By discre\-tizing the probability distribution of the distance measure, it is possible to subdivide the CUSUM sequence $S_{k}$ into a finite set of partitions. The idea is to approximate\- the continuous scheme by a Markov chain having\- $N+1$ states labeled as $\{E_{0},E_{1},\ldots,E_{N}\}$, where $E_{N}$ is absorbing. Then, the probability that the chain remains in the same state at the next step should correspond to the case when $S_{k}$ does not change in value by more than a small amount, say $\frac{1}{2}\Delta_{S}$, i.e., the next distance measure $z_{k}$ does not differ from the bias $b$ by more than $\frac{1}{2}\Delta_{S}$. The constant $\Delta_{S}$ determines the width of the grouping interval involved in the discretization of the probability distribution of $z_{k}$. The interval width $\frac{1}{2}\Delta_{S}$ must be selected such that the probability of jumping from $E_{j}$, $j \in \{0,\ldots,N-1\}$ to the absorbing state $E_{N}$ is approximately equal to the probability that the CUSUM sequence $S_{k}$ jumps beyond the threshold $\tau$ from a position $S_{k-1} \in (0,\tau)$ which corresponds approximately to the state $E_j$. This requirement is satisfied by taking
\begin{equation}\label{68}
\Delta_{S} := \frac{2\tau}{2N-1},
\end{equation}
see \cite{Evans} for details. Then, the transition probabilities from a starting state $E_j$, $j = 0,\ldots,N-1$, can be determined from the probability distribution of $z_{k} - b = r_{k}^T \Sigma^{-1} r_k-b$, as:
\begin{align*}\label{69}
&\text{pr}(E_j \rightarrow E_0) \hspace{1.9mm}= \text{pr}(z_{k} - b \leq -j\Delta_{S} + \tfrac{1}{2} \Delta_{S}),\\[1mm]
&\text{pr}(E_j \rightarrow E_{N}) \hspace{.9mm} =  \text{pr}((N-j)\Delta_{S} - \tfrac{1}{2} \Delta_{S} < z_{k} - b),\\[1mm]
&\text{pr}(E_j \rightarrow E_\nu ) \hspace{1.75mm}= \text{pr}(z_{k} - b \leq (\nu-j)\Delta_{S} + \tfrac{1}{2} \Delta_{S})\\[1mm]  &\hspace{22mm} - \text{pr}(z_{k} - b < (\nu-j)\Delta_{S} - \tfrac{1}{2} \Delta_{S}).
\end{align*}
Note that $\text{pr}(E_0 \rightarrow E_{N} ) = \text{pr}( z_{k} - b>\tau)$. For given $b$ and $\tau$, the states $\{E_0,\ldots,E_N\}$ and the above transition probabilities forms a Markov chain whose transition matrix can be constructed from the probability distribution of $z_{k}-b$. Denote $T_\chi := \text{pr}( z_{k} - b \leq \chi\Delta_{S} + \frac{1}{2} \Delta_{S})$ and $p_{\chi} := \text{pr}(\chi\Delta_{S} - \frac{1}{2} \Delta_{S} < z_{k} - b \leq \chi \Delta_{S} + \frac{1}{2} \Delta_{S})$. Then, the Markov transition matrix $\mathcal{P} \in \Real^{(N+1)\times (N+1)}$ is given by:
\begin{equation}\label{70}
\begingroup
\renewcommand*{\arraycolsep}{1.2pt}
\mathcal{P} := \begin{pmatrix}
 T_0    & p_1     & p_2          & \ldots & p_{N-1} & 1-T_{N-1} \\
 T_{-1} & p_0     & p_1          & \ldots & p_{N-2} & 1-T_{N-2}\\
 \vdots & \vdots  & \vdots       &        & \vdots  & \vdots\\
 T_{-j} &  p_{1-j}& p_{2-j}      & \ldots & p_{N-1-j} & 1-T_{N-1-j}\\
 \vdots &  \vdots & \vdots       &        & \vdots  & \vdots\\
 T_{1-N}&  p_{2-N}& p_{3-N}      & \ldots & p_{0} & 1-T_{0}\\
 0      &  0      & 0            & \ldots & 0       & 1
\end{pmatrix}.\endgroup
\end{equation}
Since the state $E_N$ is absorbing, the last row consists of zeros except for the last entry. To compute the transition probabilities $T_\chi$ and $p_{\chi}$ of $\mathcal{P}$, we need the Cumulative Distribution Function (CDF) of the shifted distance measure $z_{k} - b = r_{k}^T \Sigma^{-1} r_{k}- b$. If there are no attacks, $r_{k} \sim \mathcal{N}(\mathbf{0},\Sigma)$; therefore, $z_k- b$ follows a shifted chi-squared distribution with CDF:
\begin{equation}\label{71}
F_{z_{k}-b}(x) := \left\{
\begin{array}{ll}
 \text{P}\left(\frac{m}{2},\frac{x+b}{2} \right), \text{ \ for \ } x \geq -b,\\
0, \text{ \ for \ } x < -b,
\end{array}
\right.
\end{equation}
where $\text{P}(\cdot,\cdot)$ denotes the regularized lower incomplete gamma function \cite{Ross}. Then, the entries of the transition matrix are given by
\begin{equation}\label{72}
\left\{
\begin{array}{ll}
p_\chi = F_{z_{k}-b}\left(\chi\Delta_{S} + \tfrac{1}{2} \Delta_{S} \right)\\[1mm] \hspace{5mm} - F_{z_{k}-b}\left(\chi\Delta_{S} - \tfrac{1}{2} \Delta_{S}\right),\\[1mm]
T_\chi = F_{z_{k}-b}\left(\chi\Delta_{S} + \tfrac{1}{2} \Delta_{S}\right).
\end{array}
\right.
\end{equation}
Define the transformation $\mathcal{T}:=(I_N \hspace{1.5mm} \mathbf{0}_{N \times 1}) \in \Real^{N \times (N+1)} $ and the matrix:
\begin{equation}\label{Rm}
\mathcal{R} := \mathcal{T}\mathcal{P}\mathcal{T}^T \in \Real^{N \times N}.
\end{equation}
The matrix $\mathcal{R}$ is known as the \emph{fundamental matrix} associated with the Markov transition matrix $\mathcal{P}$. Note that
\begin{equation*}
\begingroup
\renewcommand*{\arraycolsep}{3pt}
\mathcal{P} = \begin{pmatrix}
    \mathcal{R} \hspace{6mm}   & *\\
    \mathbf{0}_{1 \times N}    & 1
\end{pmatrix}.\endgroup
\end{equation*}
Then, all entries of $\mathcal{R}$ are non-negative and its row sums are less than one. Therefore, by Gershgorin circle theorem,  the eigenvalues of $\mathcal{R}$ satisfy: $1 > |\lambda_N| \geq \ldots \geq |\lambda_1|$. It follows that $\rho[\mathcal{R}]<1$, where $\rho[\cdot]$ denotes spectral radius; therefore, the matrix $(I_N-\mathcal{R})$ is invertible \cite{Horn}. Next, having\linebreak introduced the transition matrix $\mathcal{P}$ of the approxi\-mated Markov chain and the fundamental matrix $\mathcal{R}$, we can compute an approxi\-mation $\tilde{\mathcal{A}}$ of the false alarm rate $\mathcal{A}$ based on the result in \cite{Evans}, equation (\ref{66}), and \eqref{68}-\eqref{Rm}.

\begin{theorem}\label{prop2} Assume that there are no attacks on the system\linebreak and let the \emph{CUSUM} \eqref{35} with bias  $b>\bar{b} = m$ and \linebreak threshold $\tau \in \Real_{>0}$ be driven by the distance measure \linebreak $z_{k} = r_{k}^T \Sigma^{-1} r_k$ with residual sequence $r_{k} \sim \mathcal{N}(\mathbf{0},\Sigma)$, $k \in \Nat$. For a finite number of partitions $N \in \Nat$, consider the fundamental matrix $\mathcal{R}$, defined in \eqref{Rm}, obtained from the transition matrix $\mathcal{P}$ \eqref{68}-\eqref{72}, and define
\begin{equation}\label{73}
\mu := (I_{N}-\mathcal{R})^{-1}\mathbf{1}_{N\times1} = [\mu_{1},\dots,\mu_{N}]^T.
\end{equation}
Then, the false alarm rate $\mathcal{A}=1/\emph{\text{ARL}},$ is approximately given by $\tilde{\mathcal{A}} := \mu_{1}^{-1}$. Moreover, as $N \rightarrow \infty$, $\tilde{\mathcal{A}} \rightarrow \mathcal{A}$, i.e., $\lim_{N \rightarrow \infty} \tilde{\mathcal{A}} = \mathcal{A}$.
\end{theorem}
\textbf{\emph{Proof}}: Consider the Markov chain $\mathcal{M}$ given by the states $\{E_0,E_1,\ldots,E_{N}\}$ and the transition matrix $\mathcal{P}$ \eqref{68}-\eqref{72}. Let $\tilde{\mathcal{K}} \in \Nat $ denote the number of iterations needed to reach the absorbing state $E_{N}$ from $E_{0}$. The random variable $\tilde{\mathcal{K}}$ follows a \emph{discrete phase-type distribution} with $E[\tilde{\mathcal{K}}] = \mu_{1}$ and $\mu_{1}$ as defined in (\ref{73}), see \cite{Dayar}. By construction,\linebreak $\mathcal{M}$ is a finite state approximation of the continuous Markov chain formed by the CUSUM sequence $S_{k} \in \Real_{\geq 0}$, $k \in \Nat$ driven by $z_{k} = r_{k}^T \Sigma^{-1} r_k$, $k \in \Nat$. It follows that $E[\tilde{\mathcal{K}}] = \mu_{1} \approx E[\mathcal{K}] = \text{ARL}$ where $\mathcal{K}$ denotes the \emph{run length} of the CUSUM defined in (\ref{65}). Then, from (\ref{66}), we have that $\mathcal{A} = \text{ARL}^{-1} \approx \mu_{1}^{-1}$. Next, increasing the number of partitions $N$ would reduce the width of the grouping interval $\Delta_S$ (\ref{68}), such that, as $N \rightarrow \infty$, the Markov chain $\mathcal{M}$ retrieves the continuous scheme given by the CUSUM sequence $S_k \in \Real_{\geq 0}$, $k \in \Nat$, \eqref{35}; therefore, $\mathcal{A} =  1/\lim_{N \rightarrow \infty} E[\tilde{\mathcal{K}}]$
. \hfill $\blacksquare$

\begin{remark}
Theorem 2 provides a tool for approximating the false alarm rate $\mathcal{A}$ of the \emph{CUSUM} procedure for given bias $b$ and threshold $\tau$. In particular, for a given $b>\bar{b}$, it provides a map $\mathcal{S}: \Real_{> 0} \rightarrow (0,1)$ from the threshold $\tau$ to the approximated false alarm rate $\tilde{\mathcal{A}}$, i.e., $\tau \mapsto \mathcal{S}(\tau)$, \linebreak $\tilde{\mathcal{A}} = \mathcal{S}(\tau)$. Given that $F_{z_{k}-b}(z)$ is a continuous function for all $z \in \Real$, it can be proved that $\mathcal{S}(\tau)$ is continuous for all $\tau \in \Real_{>0}$\emph{;} then, simple bisection methods can be used to determine the threshold $\tau = \tau^* \in \Real_{>0}$ required to satisfy $\tilde{\mathcal{A}} = \mathcal{S}(\tau^*) = \mathcal{A}^*$ for given $b>\bar{b}$.
\end{remark}

\section{Chi-squared Tuning}
The CUSUM approach to fault/attack detection offers an compelling alternative to the more popular chi-squared detector. Here, we use the chi-squared approach as a benchmark to compare the performance of the CUSUM. Consider again the residual sequence $r_{k}$, \eqref{26}, and its covariance matrix $\Sigma$, \eqref{28}. The chi-squared procedure is defined as follows:
\noindent\rule{\hsize}{1pt}\vspace{.2mm}
\textbf{Chi-squared procedure:}
\begin{equation}\label{baddata}
\text{If \ } z_k = r_{k}^T \Sigma^{-1} r_k  > \alpha, \hspace{2mm} \tilde{k} = k.
\end{equation}
\textbf{Design parameter:} threshold $\alpha \in \Real_{>0}$.\\
\textbf{Output:} alarm time(s) $\tilde{k}$.\\
\vspace{.2mm}\noindent\rule{\hsize}{1pt}\vspace{1mm}
The idea is that alarms are triggered if $z_{k}$ exceeds the threshold $\alpha$. Similar to the CUSUM procedure, the parameter $\alpha$ is selected to satisfy a required false alarm rate $\mathcal{A}^*$.\\

\begin{theorem}\label{prop3} Assume that there are no attacks on the system and consider the chi-squared procedure \eqref{baddata} with threshold $\alpha \in \Real_{>0}$, $r_{k} \sim \mathcal{N}(\mathbf{0},\Sigma)$. Let  $\alpha = \alpha^* := 2 \text{\emph{P}}^{-1}(\frac{m}{2},1-\mathcal{A}^*)$, where $\emph{\text{P}}^{-1}(\cdot,\cdot)$ denotes the inverse regularized lower incomplete gamma function, then $\mathcal{A} = \mathcal{A}^*$.
\end{theorem}
\textbf{\emph{Proof}}: Let $\tilde{\mathcal{K}}$ denote the {run length} of the chi-squared procedure (\ref{baddata}) defined as the number of iterations needed such that $k=\mathcal{\tilde{\mathcal{K}}}$ implies $z_{k}  > \alpha$ when there are no attacks. As with the CUSUM procedure, the {Average Run Length} is given by $\text{ARL}=E[\tilde{\mathcal{K}}]$ and the false alarm rate satisfies $\mathcal{A}=1/\text{ARL}$. The random variable $\tilde{\mathcal{K}}$ follows a geometric distribution \cite{Ross}; therefore, $\text{ARL}=E[\tilde{\mathcal{K}}]=1/\text{pr}(z_{k}>\alpha)$ and $\mathcal{A} = \text{pr}(z_{k}>\alpha)$. Each element of the sequence $z_{k}$, $k \in \Nat$ is an i.i.d. random variable with CDF given by $\tilde{F}_{z_{k}}(x) = \text{P}(\frac{m}{2},\frac{x}{2})$. Then, $\mathcal{A} = \text{pr}(z_{k}>\alpha) = 1-\text{P}(\frac{m}{2},\frac{x}{2})$ and the result follows. \hfill $\blacksquare$

\section{Detector Performance under Zero-Alarm Attacks}

In this section, we assess the performance of the CUSUM procedure by quantifying the effect of the attack sequence $\delta_{k}$ on the estimation error when the CUSUM procedure is used to identify anomalies. To maintain an equitable comparison between detectors in this section, some assumption must be made about their false positive rate (false alarm rate) and false negative rate (the rate at which true attacks are not detected). Using the tools introduced in prior sections, we can calibrate the CUSUM and chi-squared detectors to have the same false alarm rate. Here, we consider a class of \emph{zero-alarm attacks}, i.e., attack sequences that keep the detector from raising alarms. This implies that the entire attacked distribution (of $z_k$ or $S_k$) is at or below the decision threshold, effectively maximizing the false negative rate (since the true positive rate is zero, i.e., the true attack is never detected). Zero-alarm attacks provide a concise quantification of attacker impact on the system performance. In particular, we characterize the estimation error deviation due to zero-alarm attacks. This serves as a useful proxy for the capabilities of the attacker due to the detection mechanism. To do this, using the notion of \emph{input to state stability} \cite{SONTAG1995351}-\cite{JIANG2001857}, we derive upper bounds on the trajectories of the estimation error given the system dynamics, the attack sequence, and the CUSUM parameters. Furthermore, we compare the performance of the CUSUM against the chi-squared detector. In this paper, we have now characterized the rate of false alarms based on the type of detector and the system and detector parameters. We now close the loop on this analysis by identifying the impact that attackers can have while taking advantage of the detector structure.

%

\subsection{Zero-alarm Attacks}

Here, we quantify the damage that attacks may induce to the estimation error dynamics while enforcing that alarms are not raised by the detector. We assume that the attacker has perfect knowledge of the system dynamics, the Kalman filter, control inputs, measurements, and detection procedure (either CUSUM or chi-squared). It is further assumed that all the sensors can be compromised by the attacker at each time step (a worst-case scenario).

First, consider the chi-squared procedure (\ref{baddata}) and write $z_k$ in terms of the estimation error $e_k$:
\begin{equation}\label{76}
z_{k}  =(Ce_k + \eta_{k} + \delta_{k})^T\Sigma^{-1} (Ce_k + \eta_{k} + \delta_{k}).
\end{equation}
Because $e_k$ and $\eta_k$ have infinite support, to prevent $z_k$ from going beyond the threshold $\alpha$, the attack sequence $\delta_{k}$ must compensate for the term $Ce_k + \eta_{k}$. By assumption, the attacker has access to $y_{k} = Cx_k + \eta_{k}$ (real-time sensor measurements). Moreover, given its perfect knowledge of the Kalman filter, the adversary can compute the estimated output $C\hat{x}_k$ and then construct $y_k - C\hat{x}_k = Ce_k + \eta_{k}$. For a given chi-squared threshold $\alpha$, define the sequence $\bar{\delta}_k^{\alpha} := \{ \bar{\delta}_k^{\alpha} \in \Real^m | (\bar{\delta}_k^{\alpha})^T \bar{\delta}_k^{\alpha} \leq \alpha\}$;\linebreak for instance, $\bar{\delta}_k^\alpha = [\sqrt{\frac{\alpha}{m}},\sqrt{\frac{\alpha}{m}},\ldots,\sqrt{\frac{\alpha}{m}}]^T$ and $\bar{\delta}_k^\alpha = [\sqrt{\alpha},0,\ldots,0]^T$. Let $k = k^*$ denote the starting attack instant for some $k^* \geq 1$. Then, for $k \geq k^*$, it follows that
\begin{equation}\label{76}
\delta_{k} = -Ce_k - \eta_{k} + \Sigma^{\frac{1}{2}} \bar{\delta}_k^{\alpha} \rightarrow z_{k} \leq \alpha,
\end{equation}
where $\Sigma^{\frac{1}{2}}$ denotes the symmetric square root matrix of $\Sigma$, is a feasible attack sequence given the capabilities of the attacker. Sequences $\delta_k$ of the form \eqref{76} define a class of attacks that can be launched by the opponent while preventing the chi-squared detector from raising alarms, i.e., zero-alarm attacks. The estimation error dynamics under the attack (\ref{76}) is given by
\begin{equation}\label{77}
e_{k+1} = Fe_k -  L \Sigma^{\frac{1}{2}} \bar{\delta}_k^\alpha +v_k, \hspace{2mm} k \geq k^*.
\end{equation}

\begin{remark}
Note that if $\rho[F] > 1$, then $\norm{E[e_k]}$ diverges to infinity as $k$ grows for any nonstabilizing $\bar{\delta}_k^\alpha$ \emph{\cite{Astrom}}. That is, zero-alarm attacks of the form \eqref{76} may destabilize the system if $\rho[F] > 1$. If $\rho[F] \leq 1$, then $\norm{E[e_k]}$ may or may not diverge to infinity depending on algebraic and geometric multiplicities of the eigenvalues with unit modulus of $F$ (a known fact from stability of LTI systems \emph{\cite{Astrom}}).
\end{remark}

Using the superposition principle of linear systems, the estimation error $e_k$ can be written as $e_k = e_{k}^v + e_{k}^\delta$,  where $e_{k}^v$ denotes the part of $e_k$ driven by noise and $e_{k}^\delta$ is the part driven by attacks. Using this new notation, we can write the dynamics \eqref{77} as follows:
\begin{align}\label{77cc}
&e_{k+1}^v = Fe_{k}^v + v_k,\\
&e_{k+1}^\delta = Fe_{k}^\delta -  L \Sigma^{\frac{1}{2}} \bar{\delta}_k^\alpha, \hspace{2mm} k \geq k^*,\label{77dd}
\end{align}
with $e_{k^*}^v=e_{k^*}$ and $e_{k^*}^\delta=\mathbf{0}$. Therefore, the contribution of zero-alarm attacks to $e_k$ is solely determined by $e_{k}^\delta$ generated by \eqref{77dd}. For a sequence $s_k \in \Real^n$, $k \in \Nat$, let $s_{[k^*,k]}$ denote the truncation of $s_k$ from $k^*$ to $k$, i.e., $s_{[k^*,k]}:= \{s_{k^*},\ldots,s_k\}$ and $||s_{[k^*,k]}|| := \sup_{k^* \leq N \leq k} \norm{s_N}$. For any matrix $A^{n \times n}$ such that $\rho[A]<1$, let $\norm{\cdot}_*$ denote some matrix norm satisfying $\norm{A}_*<1$ (such a norm always exists if $\rho[A]<1$ \cite{Horn}).

\begin{proposition}\label{prop4}
Consider the process \eqref{17}, the Kalman filter \linebreak  \eqref{19}-\eqref{24}, and the chi-squared procedure \eqref{baddata} with threshold $\alpha \in \Real_{>0}$. Assume $\rho[F]<1$ and let $c \in \Real_{>0}$ be some constant satisfying $\norm{F^k} \leq c \norm{F^k}_*$ for all $k \in \Nat$. Let the sensors be attacked by the sequence \eqref{76}; then, for all $\bar{\delta}_{[k^*,k]}^\alpha$, $k > k^* \in \Nat$, the trajectories of \eqref{77dd} satisfy the inequalities:
\begin{align}\label{ISS1}
&\left\{
\begin{array}{lll}
\norm{e_k^{\delta}} \leq \gamma^{\chi^2}_k :=  \sqrt{\alpha} \hspace{.5mm} c ||L\Sigma^{\frac{1}{2}}|| \dfrac{ 1-\norm{F}_*^{k-k^*} }{1-\norm{F}_* \hspace{5mm}},\\[3mm]
\lim_{k \rightarrow \infty}\norm{e_k^\delta} \leq \bar{\gamma}^{\chi^2} :=   \dfrac{ \sqrt{\alpha} \hspace{.5mm} c ||L\Sigma^{\frac{1}{2}}|| }{1-\norm{F}_*}.
\end{array} \right.
\end{align}
\end{proposition}
\textbf{\emph{Proof}}: The solution of \eqref{77dd}, for $k > k^*$, is given by
\begin{align*}
e_k^\delta = - \sum_{i=0}^{k-1-k^*}F^iL\Sigma^{\frac{1}{2}}\bar{\delta}_{k-i-1}^\alpha,
\end{align*}
it follows that
\begin{equation*}
\begin{array}{lll}
\norm{e_k^\delta} \leq ||L\Sigma^{\frac{1}{2}}|| \sum_{i=0}^{k-1-k^*}\norm{F^i} ||\bar{\delta}_{[k^*,k-1]}^\alpha||, \hspace{2mm} k > k^*.
\end{array}
\end{equation*}
Because $\rho[F]<1$, there exists a matrix norm, say $\norm{\cdot}_*$, such that $\norm{F}_*<1$ (see Lemma 5.6.10 in \cite{Horn}). Moreover, because all norms are equivalent in finite dimensional vector
spaces \cite{Horn}, there exists a constant $c \in \Real_{>0}$ satisfying $\norm{D} \leq c\norm{D}_*$ for all $D \in \Real^{n \times n}$ \cite{Stewart}. It follows that
\begin{equation*}
\begin{array}{lll}
\norm{e_k^\delta} \leq c ||L\Sigma^{\frac{1}{2}}|| \sum_{i=0}^{k-1-k^*}\norm{F}_*^i ||\bar{\delta}_{[k^*,k-1]}^\alpha||\\[2mm]
\hspace{8mm} \leq c||L\Sigma^{\frac{1}{2}}|| \dfrac{ 1-\norm{F}_*^{k-k^*}}{1-\norm{F}_* \hspace{5mm}}||\bar{\delta}_{[k^*,k-1]}^\alpha||,
\end{array}
\end{equation*}
because $\sum_{i=0}^{n}\norm{F}_*^i$ is a geometric series. By construction, for all $k \geq k^*$, $(\bar{\delta}_k^\alpha)^T \bar{\delta}_k^\alpha \leq \alpha$ which implies $||\bar{\delta}_{[k^*,k-1]}^\alpha|| = \sup_{k^* \leq N \leq k-1} \norm{\bar{\delta}_{N}^\alpha} \leq \sqrt{\alpha}$; therefore, the estimation error driven by attacks $e_k^\delta$ satisfies the inequalities in \eqref{ISS1}. \hfill $\blacksquare$ \vspace{1mm}

The effect of the attack sequence \eqref{76} on the upper bound of the estimation error \eqref{ISS1} is determined by the sequence $\gamma^{\chi^2}_k$. The sequence $\gamma^{\chi^2}_k$ quantifies the impact of zero-alarm attacks when the chi-squared detector is used to detect anomalies, i.e., $\gamma^{\chi^2}_k$ gives a measure of the detector performance for the class of attacks in \eqref{76} in terms of estimation error deviation. The sequence $\gamma^{\chi^2}_k$ depends on the norm $\norm{\cdot}_{*}$ which could be any matrix norm satisfying $\norm{F}_{*}<1$. If usual matrix norms (e.g., $\norm{\cdot}_{1}$, $\norm{\cdot}_{2}$, $\norm{\cdot}_{\infty}$, etc.) do not satisfy this condition, in the proof of Lemma 5.6.10 in \cite{Horn}, the authors give a procedure for constructing such a norm provided that $\rho(F)<1$. For given norm $\norm{\cdot}_{*}$ satisfying $\norm{F}_{*}<1$, the constant $c$ can be taken as $c = \inf \{c \in \Real_{>0}:  \norm{F^k} - c\norm{F^k}_{*} \leq 0, \hspace{.5mm} \forall \hspace{.5mm} k \in \Nat \}$  which can be obtained numerically.

Next, consider the CUSUM procedure and write (\ref{35}) in terms of the estimation error $e_k$:
\begingroup\makeatletter\def\f@size{10}\check@mathfonts
\def\maketag@@@#1{\hbox{\m@th\normalsize\normalfont#1}}%
\begin{equation}\label{79}
S_{k} = \max(0,S_{k-1} +  ||\Sigma^{-\frac{1}{2}} (Ce_k + \eta_{k} + \delta_{k})||^2 - b ) ,
\end{equation}\endgroup
if $S_{k-1} \leq \tau$; and $S_{k}=0$, if $S_{k-1} > \tau$. As with the chi-squared procedure, we look for attack sequences that maintain the CUSUM statistic below the threshold $\tau$ preventing alarms to be raised. Let the attack start at some $k=k^* \geq 2$ and $S_{k^*-1} \leq \tau$, i.e., the attack does not start immediately after a false alarm. Define $\bar{\tau}_{k} := \{ \bar{\tau}_{k} \in \Real^m | \bar{\tau}^T_{k} \bar{\tau}_{k} \leq \tau + b-S_{k-1}\}$ and $\bar{\delta}_k^b := \{ \bar{\delta}_k^b \in \Real^m | (\bar{\delta}_k^b)^T \bar{\delta}_k^b \leq b\}$ for given threshold $\tau$ and bias $b$. Consider the attack sequence:
\begin{align}\label{80}
\delta_{k} = \left\{\begin{array}{ll}
 -Ce_k - \eta_{k} + \Sigma^{\frac{1}{2}}\bar{\tau}_{k}, \hspace{2mm} k = k^*, \\
 -Ce_k - \eta_{k} + \Sigma^{\frac{1}{2}}\bar{\delta}_k^b, \hspace{2mm}k > k^*.
\end{array} \right.
\end{align}
It follows that $S_{k^*} = \max(0,S_{k^*-1} +  \bar{\tau}^T_{k} \bar{\tau}_{k} - b ) \leq \tau$, $S_{k^*+1} = \max(0,S_{k^*} +  (\bar{\delta}_{k^*+1}^b)^T \bar{\delta}_{k^*+1}^b - b ) \leq \tau$, and $S_{k^*+N} = \max(0,S_{k^*+N-1} +  (\bar{\delta}_{k^*+1}^b)^T \bar{\delta}_{k^*+1}^b - b ) \leq \tau$ for all $N \in \Nat$. That is, the class of attack sequences in \eqref{80} prevents the CUSUM procedure from raising alarms. Note that the attacker can only induce this sequence by exactly knowing $S_{k^*-1}$, i.e., the value of the CUSUM sequence one step before the attack. This is a strong assump\-tion since it represents a real-time quantity that is not commu\-nicated over the communication network. Even if the opponent has access to the parameters of the CUSUM $(b,\tau)$ given the stochastic nature of the residual, the attacker would need to know the complete history of observations (from when the CUSUM was started) to be able to reconstruct $S_{k^*-1}$ from data. This is an inherent security advantage in favor of the CUSUM over static detectors like the bad-data or chi-squared. Nevertheless, for evaluating the worst case scenario, we assume that the attacker has access to $S_{k^*-1}$. By construction, the estimation error dynamics under the attack sequence (\ref{80}) is written as: $e_{k^*+1} = Fe_k^* - L\Sigma^{\frac{1}{2}}\bar{\tau}_{k^*} + v_{k^*}$, and, for $k > k^*$,
\begin{equation}\label{77bb}
e_{k+1} = Fe_k -  L \Sigma^{\frac{1}{2}} \bar{\delta}^b_k + v_k.
\end{equation}
Note that \eqref{77} and \eqref{77bb} have the same dynamics but different initial condition. Therefore, we may expect upper bounds on $\norm{e_k}$ similar to \eqref{ISS1} obtained for the chi-squared. Again, we write $e_k$ as $e_k = e_{k}^v + e_{k}^\delta$,  where $e_{k}^v$ denotes the part of $e_k$ driven by noise and $e_{k}^\delta$ is the part driven by attacks of \eqref{77bb}. Using this new notation, we can write the dynamics \eqref{77bb} as:
\begin{align}\label{77gg}
&e_{k+1}^v = Fe_{k}^v + v_k,\\
&e_{k+1}^\delta = Fe_{k}^\delta -  L \Sigma^{\frac{1}{2}} \bar{\delta}_k^b, \hspace{2mm} k > k^*,\label{77ff}
\end{align}
with $e_{k^*}^\delta=\mathbf{0}$, $e_{k^*+1}^\delta=-L\Sigma^{\frac{1}{2}}\bar{\tau}_{k^*}$, $e_{k^*}^v=e_{k^*}$, and $e_{k^*+1}^v=Fe_{k^*}+v_{k^*}$.

\begin{proposition}\label{prop5}
Consider the process \eqref{17}, the Kalman filter \linebreak \eqref{19}-\eqref{24}, and the CUSUM procedure \eqref{35} with threshold \linebreak$\tau \in \Real_{>0}$ and bias $b > \bar{b} = m \in \Nat_{>0}$. Assume $\rho[F]<1$ and let $c \in \Real_{>0}$ be some constant satisfy $\norm{F^k} \leq c \norm{F^k}_*$ for all $k \in \Nat$. Let the sensors be attacked by the sequence \eqref{80}; then, for all $\bar{\tau}_{k^*}$ and $\bar{\delta}_{[k^*,k]}^b$, $k > k^* \in \Nat$, the trajectories of \eqref{77ff} satisfy the inequalities:
\begin{align}\label{ISS221}
&\left\{
\begin{array}{lll}
\norm{e_k^{\delta}} \leq \gamma^{\text{\emph{CS}}}_k :=  \sqrt{b} \hspace{.5mm} c ||L\Sigma^{\frac{1}{2}}|| \dfrac{ 1-\norm{F}_*^{k-k^*} }{1-\norm{F}_* \hspace{5mm}}\\[3mm] \hspace{23mm} + \hspace{.5mm} c||L\Sigma^{\frac{1}{2}}\bar{\tau}_{k^*}||\norm{F}^{k-k^*-1}_*,\\[2mm]
\lim_{k \rightarrow \infty}\norm{e_k^\delta} \leq \bar{\gamma}^{\text{\emph{CS}}} :=   \dfrac{ \sqrt{b} \hspace{.5mm} c ||L\Sigma^{\frac{1}{2}}|| }{1-\norm{F}_*}.
\end{array} \right.
\end{align}
\end{proposition}
\textbf{\emph{Proof}}: The solution of \eqref{77ff}, for $k > k^*+1$, is given by
\begin{align*}
e_k^{\delta} &= - F^{k-k^*-1}L\Sigma^{\frac{1}{2}}\bar{\tau}_{k^*} - \sum_{i=0}^{k-2-k^*} F^iL\Sigma^{\frac{1}{2}}\bar{\delta}_{k-i-1}^b.
\end{align*}
Then, following the same lines as in the proof of Proposition 1, we can write the following
\begin{align*}
\norm{e_k^\delta} &\leq  c||L\Sigma^{\frac{1}{2}}\bar{\tau}_{k^*}||\norm{F}^{k-k^*-1}_*\\ &+ c||L\Sigma^{\frac{1}{2}}||\dfrac{1-\norm{F}_*^{k-k^*} }{1-\norm{F}_*\hspace{5mm}}||\bar{\delta}_{[k^*+1,k-1]}^b||,
\end{align*}
By construction, for all $k > k^*$, $(\bar{\delta}_k^b)^T \bar{\delta}_k^b \leq b$ which implies\linebreak $||\bar{\delta}_{[k^*+1,k-1]}^b|| = \sup_{k^*+1 \leq N \leq k-1} \norm{\bar{\delta}_{N}^b} \leq \sqrt{b}$; therefore, the trajectories of the estimation error dynamics \eqref{77ff} satisfy the inequalities in \eqref{ISS221}. \hfill $\blacksquare$\vspace{1mm}

When considering the CUSUM, the effect of attacks of the form \eqref{80} on the upper bound of the estimation error is determined by the sequence $\gamma^{\text{CS}}_k$ in \eqref{ISS221}. Note that, in steady state, the ($\bar{\tau}_{k^*}$)-dependent term in $\gamma^{\text{CS}}_k$ decreases exponentially to zero. Then, the sequences $\gamma^{\chi^2}_k$ and $\gamma^{\text{CS}}_k$ take constant values asymptotically. It follows that we can directly quantify the performance ratio (in terms of steady state deviation of $\norm{e_k^\delta}$) between the two detectors. This is stated in the following corollary of Proposition 1 and Proposition 2.

\begin{corollary}
In steady state:
\begin{align}\label{ststate}
\lim_{k \rightarrow \infty} \frac{\gamma^{\chi^2}_k}{\gamma^{\text{\emph{CS}}}_k}  = \frac{\bar{\gamma}^{\chi^2}}{\bar{\gamma}^{\text{\emph{CS}}}} =  \sqrt{ \frac{\alpha}{b}}.
\end{align}
\end{corollary}

\subsection{Detector Comparison}\label{Diss}

For the case of zero-alarm attacks, we have derived upper bounds on $e_k^\delta$ for both the chi-squared and CUSUM procedures provided that $\rho[F]<1$ (otherwise $\norm{e_k^\delta}$ diverges under the class of zero-alarm attacks considered). To compare these results, we use the ratio of sequences $(\gamma^{\chi^2}_k/\gamma^{\text{\emph{CS}}}_k)$. From Coro\-llary 1, $\lim_{k \rightarrow \infty} (\gamma^{\chi^2}_k/\gamma^{\text{CS}}_k) = \sqrt{\alpha/b}$. That is, in steady state, the ($\bar{\tau}_{k^*}$)-dependent term in $\gamma^{\text{CS}}_k$ exponentially decreases to zero and, if $b<\alpha$, under the same class of zero-alarm attacks, the CUSUM procedure leads to smaller steady state deviations on $\norm{e_k}$ than the chi-squared procedure. In general, to increase the chances of attack detection, it is desired to select $b$ as close as possible to $\bar{b}$ in Theorem \ref{prop1}. It follows that $b \approx \bar{b} = m$. On the other hand, according to Theorem \ref{prop3}, $\alpha$ must be selected as $\alpha = \alpha^* = 2\text{P}^{-1}(\frac{m}{2},1-\mathcal{A}^*)$ to fulfill a desired false alarm rate $\mathcal{A}^*$. In this case, we want to select $\mathcal{A}^*$ close to zero, such that there are only a few false alarms. Let $\mathcal{A}^* \in \{0.01,0.1\}$ and $m=2$, i.e., false alarms bet\-ween $1\%$ and $10\%$ and two dimensional outputs; then, $\alpha = 2\text{P}^{-1}(\frac{m}{2},1-\mathcal{A}^*) \in [4.60,9.21]$ and $b \approx \bar{b} = 2$. This implies that, in steady state, for the same class of attacks and $\mathcal{A}^* \in [0.01,0.1]$, the chi-squared procedure leads to at least two times larger upper bounds than the CUSUM. Actually, for having $\alpha=b$ (which implies $\lim_{k \rightarrow \infty} (\gamma^{\chi^2}_k/\gamma^{\text{CS}}_k) = 1$), it is necessary to allow for a rate of $\mathcal{A}^* = 0.63$, which is high for practical purposes. For the CUSUM procedure, the thres\-hold $\tau$ is selected to fulfill the desired $\mathcal{A}^*$. Given that there are no exact closed-form expressions to relate $\tau$ and $\mathcal{A}^*$ (we provided a numeric approximation), it is not possible to exactly tell how large the $\tau$ would need be to sa\-tisfy $\mathcal{A}^*$. However, as mentioned, the contribution of $\bar{\tau}_{k^*}$ to $e_k$ vanishes exponentially, i.e., independent of how large $\tau$ is, its contribution to $\gamma^{\text{CS}}_k$ is zero in steady state.

\begin{figure}[t]
  \centering
  \includegraphics[scale=.175]{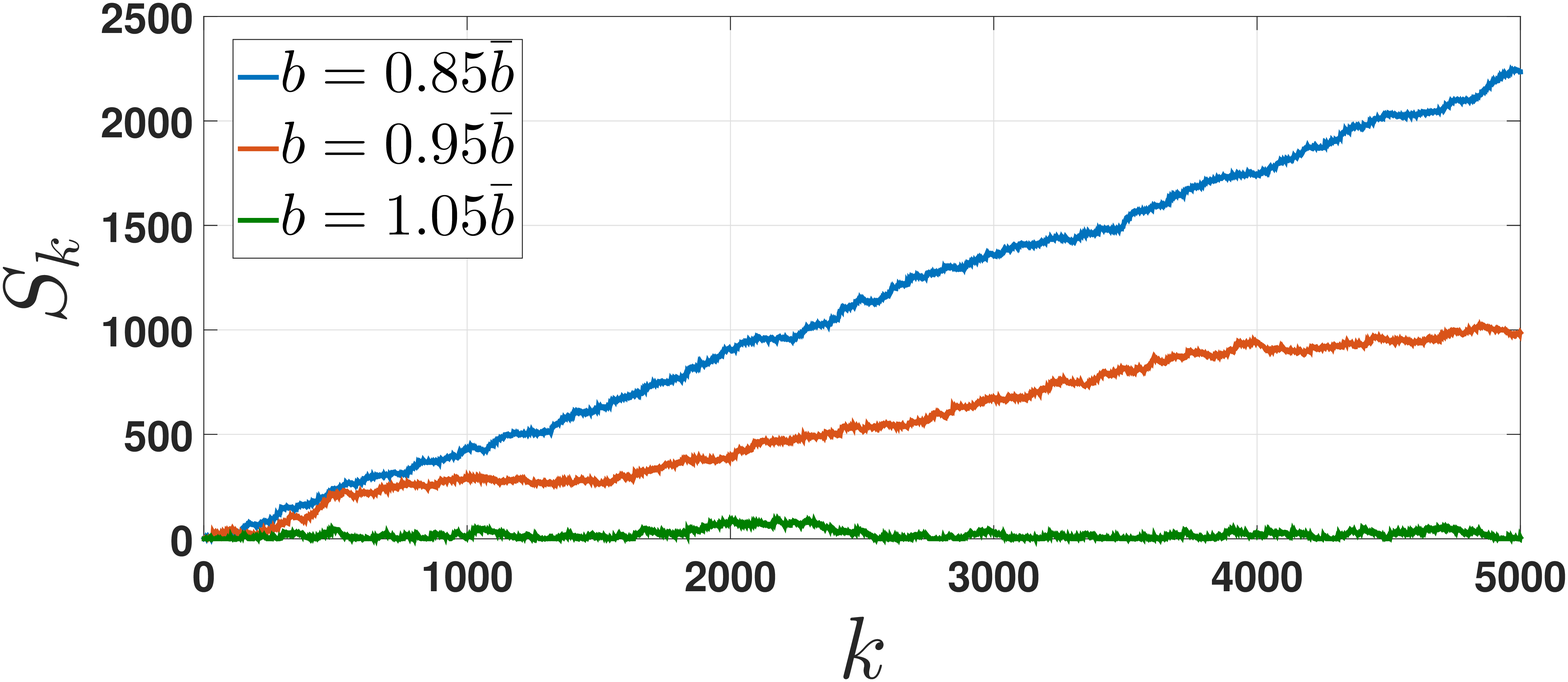}
  \caption{CUSUM evolution for different values of bias $b$.}\label{Fig2}
\end{figure}

%

\begin{figure}[t]
  \centering
  \includegraphics[scale=.179]{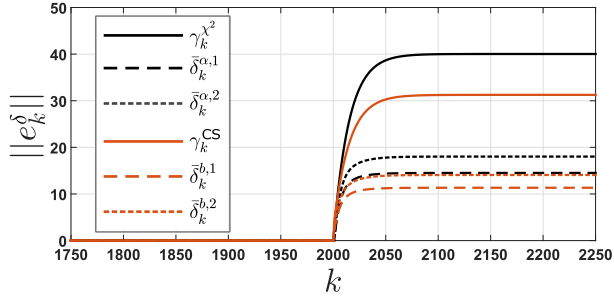}
  \caption{Upper bounds $\gamma^{\chi^2}_k$ and $\gamma^{\text{CS}}_k$, and deviation of $\norm{e_k^\delta}$ due to the zero-alarm attacks $(\bar{\delta}^{\alpha,1}_k,\bar{\delta}^{b,1}_k)$ and $(\bar{\delta}^{\alpha,2}_k,\bar{\delta}^{b,2}_k)$. Attacks are induced at $k=2 \times 10^3$.}\label{Fig4d}
\end{figure}

\begin{figure}[t]
  \centering
  \includegraphics[scale=.175]{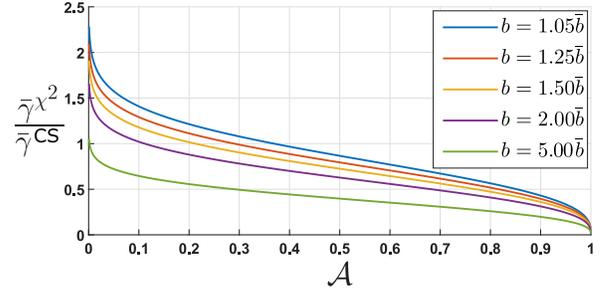}
  \caption{Asymptotic ratio $(\bar{\gamma}^{\chi^2}/\bar{\gamma}^{\text{CS}})$ versus the false alarm rate $\mathcal{A}$ for different values of CUSUM bias $b$.}\label{Fig5c}
\end{figure}

\section{Simulation Experiments}

\begin{table*}[t]
\centering
\label{table_1}
\begin{tabular}{|c|c|c|c|c|c|c|}
\hline
                             & \multicolumn{2}{c|}{$\mathcal{A}^*=0.25$}                                     & \multicolumn{2}{c|}{$\mathcal{A}^*=0.10$}                                    & \multicolumn{2}{c|}{$\mathcal{A}^*=0.02$}                                                                              \\ \hline
\multirow{2}{*}{$b/\bar{b}$} & \multirow{2}{*}{$\tau = \tau^*$} & \multirow{2}{*}{$\mathcal{A}$  (Simul.)} & \multirow{2}{*}{$\tau = \tau^*$ } & \multirow{2}{*}{$\mathcal{A}$ (Simul.)} & \multicolumn{1}{l|}{\multirow{2}{*}{$\tau = \tau^*$}} & \multicolumn{1}{l|}{\multirow{2}{*}{$\mathcal{A}$ (Simul.)}} \\
                             &                                    &                                          &                                    &                                         & \multicolumn{1}{l|}{}                                   & \multicolumn{1}{l|}{}                                        \\ \hline
$1.05$                       & $1.0282$                   & $0.2041$                                 & $3.9602$                           & $0.0899$                                & $12.3208$                                                & $0.0196$                                                     \\ \hline
$1.15$                       & $0.6872$                   & $0.2010$                                 & $3.3699$                           & $0.0885$                                & $10.0327$                                                & $0.0184$                                                     \\ \hline
$2.00$                       & $-$                           & $-$                                 & $0.2528$                           & $0.0953$                                & $4.1002$                                                & $0.0202$                                                     \\ \hline
\end{tabular}\\[4mm]
Table 1. Simulation Experiments. Results from Theorem 2 and Remark 1.
\end{table*}

The authors in \cite{Patton_Book},\cite{Wata} study the fault detection problem for a well stirred chemical reactor with heat exchanger. We use this system to demonstrate our results. The state, input, and output vectors of the considered reactor are:
\begingroup\makeatletter\def\f@size{9.0}\check@mathfonts
\def\maketag@@@#1{\hbox{\m@th\normalsize\normalfont#1}}%
\begin{align*}
\left\{
\begin{array}{ll}
x(t) :=  \begin{pmatrix} C_0\\T_0\\T_w\\T_m \end{pmatrix},
u(t) :=  \begin{pmatrix} C_u\\T_u\\T_{w,u} \end{pmatrix},
y(t) :=  \begin{pmatrix} C_0\\T_0\\T_{w} \end{pmatrix},
\end{array}
\right.
\end{align*}\endgroup
where
\begingroup\makeatletter\def\f@size{9.0}\check@mathfonts
\def\maketag@@@#1{\hbox{\m@th\normalsize\normalfont#1}}%
\begin{align*}
\left\{
\begin{array}{ll}
C_0&: \text{Concentration of the chemical product},\\
T_0&: \text{Temperature of the product},\\
T_w&: \text{Temperature of the jacket water of heat exchanger},\\
T_m&: \text{Coolant temperature},\\
C_u&: \text{Inlet concentration of reactant},\\
T_u&: \text{Inlet temperature},\\
T_{w,u}&: \text{Coolant water inlet temperature}.
\end{array}
\right.
\end{align*}\endgroup
We linearize the nonlinear model introduced in \cite{Patton_Book} about the origin $x(t) = \mathbf{0}_{4 \times 1}$ and then discretize it with sampling time $h=0.05$. The resulting discrete-time linear system is given by \eqref{17}-\eqref{24} with matrices as given in \eqref{Simul}. The original model in \cite{Patton_Book} does not consider sensor/actuator noise, we have included noise to increase the complexity of our simu\-lation experiments. First, assume\- no attacks, i.e., $\delta_{k} = \mathbf{0}$, and consider the CUSUM procedure (\ref{35}) with distance measure $z_{k}=r_{k}^T \Sigma^{-1} r_k$ and residual sequence \eqref{26}. According to Theorem \ref{prop1}, the bias $b$ must be selected larger than $\bar{b}=m=3$ to ensure mean square boundedness of $S_{k}$ independent of the threshold\- $\tau$. Figure \ref{Fig2} depicts the evolution of the CUSUM for $b \in \{0.85\bar{b},0.95\bar{b},1.05\bar{b}_1 \}$ and $k\in [1,5000]$. For the purpose of illustrating this unbounded growth, we have omitted the reset procedure of the CUSUM. Note that the bound for $b$ is tight, small deviations from $\bar{b}$ lead to (bounded\-ness) unboundedness of $S_k$. Next, for desired false alarm rates $\mathcal{A}^* \in \{ 0.25,0.10,0.02\}$, we compute the corresponding thresholds $\tau =\tau^*$ using Theorem \ref{prop2} and Remark 4. For these thresholds, in Table 1, we present the actual false alarm rate $\mathcal{A}$ (obtained by simulation) and the desired $\mathcal{A}^*$. Note that the difference between $\mathcal{A}$ and $\mathcal{A}^*$ is less that $0.05$ in all cases.

In Figure \ref{Fig4d}, we present the evolution of $\norm{e_k^\delta}$ when both the chi-squared and the CUSUM are deployed for attack\- detection and the attack sequences $\delta_{k}$ are zero-alarm attacks of the form introduced in \eqref{76} and \eqref{80}, respectively. We consider two attack sequences, first, $\bar{\delta}^{\alpha}_k = \bar{\delta}^{\alpha,1}_k=\sqrt{\alpha/m} \bar{\delta}_1$, $\bar{\delta}^{b}_k = \bar{\delta}^{b,1}_k = \sqrt{b/m}\bar{\delta}_1$, $\bar{\tau}_{k^*} = \bar{\tau}^1_{k^*}= \sqrt{(\tau + b - S_{k-1})/m} \bar{\delta}_1$, and $\bar{\delta}_1 = \mathbf{1}_{m \times 1}$. The second attack is $\bar{\delta}^{\alpha}_k = \bar{\delta}^{\alpha,2}_k=\sqrt{\alpha} \bar{\delta}_2$, $\bar{\delta}^{b}_k = \bar{\delta}^{b,2}_k = \sqrt{b}\bar{\delta}_2$, and $\bar{\tau}_{k^*} = \bar{\tau}^2_{k^*}= \sqrt{\tau + b - S_{k-1}}\bar{\delta}_2$, where $\bar{\delta}_2$ denotes the unitary singular vector corresponding to the largest singular value of $(I-F)^{-1}L\Sigma^{\frac{1}{2}}$. It can be proved that this selection of $\bar{\delta}_2$ maximizes the steady state value of $\norm{e_k^\delta}$. For the CUSUM, we select $b=2\bar{b}=6$ and $\tau=\tau^*=4.1002$ such that $\mathcal{A} \approx \mathcal{A}^* = 0.02$ (see Table 1). Likewise, we select $\alpha = \alpha^* = 2 \text{P}^{-1}(\frac{2}{2},1-0.02)=9.83$ such that, according to Theorem 3, $\mathcal{A} = \mathcal{A}^* = 0.02$. The attacks are induced at $k=k^*=2 \times 10^3$. Note that, as stated in Proposition 1 and Proposition 2, given that $\rho[F]<1$, inequalities \eqref{ISS1} and \eqref{ISS221} are satisfied for both attacks. Moreover, as mentioned in Section \ref{Diss}, we expect that the CUSUM leads to smaller steady state deviation on $\norm{e_k}$ because $b<\alpha$ and $\bar{\delta}^{\alpha,i}_k = a\bar{\delta}^{b,i}_k$, $i=1,2$ for some $a \in \Real_{>0}$. This is exactly what we see in Figure \ref{Fig4d}. Note that the ratio $\bar{\gamma}_{\chi^2}/\bar{\gamma}_{\text{CS}}=1.28$ is fixed by our choice of false alarm rate and bias. In Figure \ref{Fig5c}, we depict the evolution of the ratio $\bar{\gamma}_{\chi^2}/\bar{\gamma}_{\text{CS}}$ versus the false alarm rate $\mathcal{A}$ for different values of CUSUM bias $b$.

\section{Conclusions}

In this paper, for a class of stochastic linear time-invariant systems, we have characterized a model-based CUSUM procedure for identifying compromised sensors. In particular, steady state Kalman filters have been proposed to estimate the state of the physical process; then, these estimates have been used to construct residual variables (between sensor measurements and estimations) which drive the CUSUM procedure. Using stability results for stochastic systems and Markov chain approximations of the CUSUM sequence, we have derived systematic tools for tuning the CUSUM procedure such that mean square boundedness of the CUSUM sequence is guaranteed and the desired false alarm rate is fulfilled. For a class of zero-alarm attacks, we have characterized the performance of the proposed CUSUM procedure in terms of the effect that the attack sequence can induce on the system dynamics. Then, we have compared this performance against the one obtained using chi-squared procedures. For the linearized model of the chemical reactor considered in \cite{Wata,Patton_Book}, by means of a simulation study, we have showed that our tools are useful for tuning the CUSUM procedure and provide accurate predictions about the performance of the detection scheme.

\begin{strip}
\noindent\rule{\hsize}{1pt}
\begingroup\makeatletter\def\f@size{8.0}\check@mathfonts
\def\maketag@@@#1{\hbox{\m@th\normalsize\normalfont#1}}%
\begin{align}\label{Simul}
\left\{\begin{array}{ll}
F &=\begin{pmatrix}  0.8353 & 0 & 0 & 0 \\ 0 & 0.8324 & 0 & 0.0031 \\ 0 & 0.0001 & 0.1633 & 0\\ 0 & 0.0280 & 0.0172 & 0.9320 \end{pmatrix} \hspace{.5mm} G  = \begin{pmatrix} 0.0458 & 0 & 0 \\ 0 & 0.0457 & 0 \\ 0 & 0 & 0.0231 \\ 0 & 0.0007 & 0.0006 \end{pmatrix}, \hspace{.5mm} C = \begin{pmatrix} 1&0&0&0\\0&1&0&0\\0&0&1&0 \end{pmatrix},\\[10mm] L &= \begin{pmatrix} 0.8271 & 0 & 0 \\ 0 & 0.8243 & 0.0002 \\ 0 & 0.0002 & 0.1619 \\ 0 & 0.0481 & 0.0543 \end{pmatrix}, \hspace{.5mm} R_2 = 0.01\times I_3, \hspace{.5mm} R_0 = R_1 = I_4, \hspace{.5mm} \Sigma = \begin{pmatrix} 1.0169 & 0 & 0 \\ 0 & 1.0169 & 0.0001 \\ 0 & 0.0001 & 1.0105 \end{pmatrix}.
\end{array} \right.
\end{align}\endgroup
\noindent\rule{\hsize}{1pt}\\
\end{strip}

%
\bibliographystyle{IEEEtran}
\bibliography{autosam}

%
%
%
%
%
%
%

\appendices
\section{Proof of Theorem 1}
Define the functions $V_{k} := S_{k}^2$ and $\Delta V_{k} := E_{S_{k}}\big[ V_{k+1} \big] - V_{k}$. Along \eqref{35} with distance measure $z_{k}$ and independent of $\tau$, we have that
\begin{align}
\Delta V_{k} = E_{S_{k}}\big[ (S_{{k}} + z_{k+1} - b)^{+2}  \big] - S_{k}^2,
\label{44}
\end{align}
where $\rho^{+2}:=\max(0,\rho)^2$. Note that, by construction, $S_k,z_{k} \in \Real_{\geq 0}$ for all $k \in \Nat$. First, consider $S_k \in [b,\infty)$, it follows that
\[
(S_{{k}} + z - b)^{+2} = (S_{{k}} + z - b)^2,
\]
for all $z \in \Real_{\geq 0}$; therefore
\begin{align}\label{46a}
E_{S_{k}}\big[ V_{k+1} \big] &= E_{S_{k}}\big[ (S_{{k}} + z_{k+1} - b )^{2}],
\end{align}
for $S_k \in [b,\infty)$ and $z_{k+1} \in \Real_{\geq 0}$. Next, consider $S_{k} \in [0,b)$ which implies $S_{k}-b<0$, then
\begin{align*}
&z \in [0,b-S_{k}]  \Rightarrow \big(S_{{k}} + z - b\big)^{+2}= 0, \notag \\
&z \in (b-S_{k},\infty] \Rightarrow  \big(S_{{k}} + z - b\big)^{+2}=\big(S_{{k}} + z - b\big)^{2}.
\end{align*}
It follows that
\begin{align}\label{46b}
E_{S_{k}}\big[ V_{k+1} \big] &\leq E_{S_{k}}\big[ (S_{{k}} + z_{k+1} - b )^{2}],
\end{align}
for $S_k \in [0,b)$ and $z_{k+1} \in \Real_{\geq 0}$. Using (\ref{46a}), (\ref{46b}), and independence between $S_k$ and $z_{k+1}$, we have that
\begin{align}\label{46c}
E_{S_{k}}\big[ V_{k+1} \big] &\leq E_{S_{k}}\big[ (S_{{k}} + z_{k+1} - b )^{2}]\\[1mm]
&= (S_k-b)^2 + 2(S_k-b)E[z_{k+1}] + E[z_{k+1}^2], \notag
\end{align}
for $S_k,z_{k+1} \in \Real_{\geq 0}$. Using (\ref{38}) and the relation:
\[
\text{var}[z_k] = E[z_k^2] - E[z_k]^2,
\]
inequality \eqref{46c} amounts to
\begin{align}\label{46d}
E_{S_{k}}\big[ V_{k+1} \big] &\leq (S_k-b)^2 + 2(S_k-b+1)m + m^2,
\end{align}
and, therefore,
\begin{align}
\Delta V_{k} &= E_{S_{k}}\big[ V_{k+1}  \big] - S_{k}^2, \notag\\[1mm]
&\leq -2(b - m)S_k + (b - m)^2  + 2m,
\label{52}
\end{align}
for $S_k,z_{k+1} \in \Real_{\geq 0}$. From \eqref{52}, given that $b \in \Real_{>0}$ and $S_k \in \Real_{>0}$ by construction, it is easy to verify that
\begin{align}\label{53}
\Delta V_{k}< 0 \Leftrightarrow b > \bar{b} := m \text{ and } S_{k} > \bar{S}, \text{ } \bar{S}:= \tfrac{(b-m)^2+2m}{2(b-m)}.
\end{align}
Therefore, $b \in (\bar{b},\infty)$ implies:
\begin{equation}
\left\{
\begin{array}{ll}
\Delta V_{k} < 0 \text{ \ for}\hspace{2mm} S_{k} \in (\bar{S},\infty), \\[.5mm]
\Delta V_{k} \geq 0 \text{ \ for}\hspace{2mm} S_{k} \in [0,\bar{S}]\label{55C}.
\end{array}
\right.
\end{equation}
Recall that $\Delta V_{k} := E_{S_{k}}\big[ V_{k+1} \big] - V_{k}$. Assume that for some $k=k^* \in \Nat$, $S_{{k^*}} \in (\bar{S},\infty)$; then, from (\ref{55C}), $\Delta V_{{k^*}} < 0$ and consequently
\begin{equation}\label{56}
E_{S_{{k^*}}}[V_{k^*+1}]<V_{{k^*}}.
\end{equation}
Next, for $k=k^*+1$, let $S_{k^*+1} \in (\bar{S},\infty)$, then
\begin{equation}\label{57}
E_{S_{k^*+1}}[V_{k^*+2}]<V_{k^*+1}.
\end{equation}
Using (\ref{56}), (\ref{57}), and the property
\begin{equation}\label{58}
E_{S_{{k^*}}}[V_{k^*+2}] = E_{S_{{k^*}}}[E_{S_{k^*+1}}[V_{k^*+2}]],
\end{equation}
we have
\begin{equation}\label{59}
E_{S_{{k^*}}}[V_{k^*+2}] < E_{S_{{k^*}}}[V_{k^*+1}]<V_{{k^*}}.
\end{equation}
Continuing this way, we obtain
\begin{equation}\label{65P}
V_{{k^*}} > E_{S_{{k^*}}}[V_{k^*+1}] > \ldots > E_{S_{{k^*}}}[V_{k^*+n}],
\end{equation}
for $n \geq 2$ and $k = \{  k^*,k^*+1,\ldots, k^*+n \}$ such that $S_{{k}} \in (\bar{S},\infty)$. Therefore, from \eqref{65P}, it can be concluded that the second moment $E_{S_{{k^*}}}[V_{k}] = E_{S_{{k^*}}}[S_{{k}}^2]$ decreases monotonically for $S_{{k}} \in (\bar{S},\infty)$ and $b \in (\bar{b},\infty)$; and consequently, $E_{S_{{k^*}}}[S_{{k}}^2] < \infty$. Next, assume that for some $k=k^* \in \Nat$, $S_{{k^*}} \in [0,\bar{S}]$ and $b \in (\bar{b},\infty)$; then, from (\ref{55C}), $\Delta V_{{k^*}} \geq 0$ and, by (\ref{52}), it follows that
\begin{equation}\label{52BB}
\Delta V_{{k^*}} \leq -2\big( b - m  \big)S_{{k^*}}  + (b - m)^2  + 2m \geq 0.
\end{equation}
Since $S_{{k^*}} \in [0,\bar{S}]$, there always exist constants $a \in (0,1)$ and $\beta \in \Real_{>0}$ satisfying
\begin{align}\label{53BB}
\Delta V_{{k^*}} & \leq -2\big( b - m  \big)S_{{k^*}}  + (b - m)^2  + 2m \notag\\ &\leq -a V_{k^*} + \beta, \hspace{2.5mm}\text{for}\hspace{1mm} S_{{k^*}} \in [0,\bar{S}].
\end{align}
Given that $\Delta V_{k} = E_{S_{k}}\big[ V_{k+1} \big] - V_{k}$, by (\ref{53BB}), we have
\begin{equation}\label{54BB}
E_{S_{{k^*}}}[V_{k^*+1}] \leq (1-a)V_{{k^*}}+\beta, \hspace{2mm}\text{for}\hspace{1mm} S_{{k^*}} \in [0,\bar{S}].
\end{equation}
Next, for $k=k^*+1$, let $S_{k^*+1} \in [0,\bar{S}]$, then
\begin{equation}\label{55BB}
E_{S_{k^*+1}}[V_{k^*+2}] \leq (1-a)V_{k^*+1}+\beta.
\end{equation}
By (\ref{54BB}), (\ref{55BB}), and the property
\begin{equation}\label{58BB}
E_{S_{{k^*}}}[V_{k^*+2}] = E_{S_{{k^*}}}[E_{S_{k^*+1}}[V_{k^*+2}]],
\end{equation}
we have
\begin{eqnarray}\label{59BB}
  E_{S_{{k^*}}}[V_{k^*+2}] &\leq& E_{S_{{k^*}}}[(1-a)V_{k^*+1}+\beta] \notag\\
  &=& (1-a)E_{S_{{k^*}}}[V_{k^*+1}] +\beta \notag\\
  &\leq& (1-a)^2 V_{{k^*}}+ (1-a)\beta +\beta \label{59BB}.
\end{eqnarray}
Continuing this way, we obtain
\begin{align}
  E_{S_{{k^*}}}[V_{k^*+n}] \leq (1-a)^n V_{{k^*}} + \beta \text{$\sum\limits_{i=0}^{n-1}$}(1-a)^i,\label{65BB}
\end{align}
for $n \geq 1$ and $k = \{  k^*,k^*+1,\ldots, k^*+n \}$ such that $S_{{k}} \in [0,\bar{S}]$. Given that $a \in (0,1)$ and $\sum_{i=0}^{k-1}(1-a)^i \leq \sum_{i=0}^{\infty}(1-a)^i = \frac{1}{a}$, then, as $n \rightarrow \infty$, it is satisfied that
\begin{align}
E_{S_{{k^*}}}[S_{\infty}^2] \leq \frac{\beta}{a}.\label{66BB}
\end{align}
Therefore, by (\ref{65BB}) and (\ref{66BB}), the second moment $E_{S_{{k^*}}}[S_{{k}}^2]$ does not grow unbounded for $S_{{k}} \in [0,\bar{S}]$ and $b \in (\bar{b},\infty)$, i.e., $E_{S_{{k^*}}}[S_{{k}}^2] < \infty$. So far, combining the preliminary results presented above, we have proved that for all $S_{k} \in [0,b] \cup (b,\infty)$, the second moment is finite and either decreasing or uniformly bounded in $k \in \Nat$ provided that the conditions of Theorem 1 are satisfied. Hence, for the resi\-dual sequence $r_{k} \sim \mathcal{N}(\mathbf{0},\Sigma)$, $b > \bar{b} \rightarrow E_{S_{1}}[S_{{k}}^2] < \infty$ for all $k \in \Nat$. \hfill $\blacksquare$
\end{document}